\documentstyle[11pt,preprint,tighten,epsf,rotating,aps]{revtex}

\begin{document}
\thispagestyle{empty}
\preprint{CEBAF-TH-95-02, ISN 95-100}
\title{Application of HQET to $B\to K^{(*)}$ Transitions}
\author{
W. Roberts\footnote{National Young Investigator}}
\address{
Institut des Sciences Nucl\'eaires\\
53 avenue des Martyrs, 38026 Grenoble, France\\
and\\
Department of Physics, Old Dominion University, Norfolk, VA 23529,
USA \\
and \\
Continuous Electron Beam Accelerator Facility \\
12000 Jefferson Avenue, Newport News, VA 23606, USA.}
\author{F. Ledroit}
\address{
Institut des Sciences Nucl\'eaires\\
53 avenue des Martyrs, 38026 Grenoble, France.}
\maketitle
\begin{abstract}
We examine the measured rates for the decays $D\to K^{(*)}\ell\nu$,
$B\to K^{(*)}\psi^{(\prime)}$ and $B\to K^*\gamma$ in a number of
scenarios, in the framework of the heavy
quark effective theory. We attempt to find a scenario in which all of
these decays are described by a single set of
form factors. Once such a scenario is found, we make predictions for
the rare decays $B\to K^{(*)}\ell^+\ell^-$.
While we find that many scenarios can provide adequate descriptions
of all the data, somewhat surprisingly, we observe that two popular
choices of form factors,
namely monopolar forms and exponential forms, exhibit some
shortcomings, especially when confronted with polarization
observables. We predict $Br(\bar B^0\to \bar K^0\mu^+\mu^-)=
6.4\pm 1.0\times 10^{-7}$ and $Br(\bar B^0\to \bar K^{*0}\mu^+\mu^-)=
3.8\pm 1.3\times 10^{-6}$. We also make predictions for polarization
observables in these decays.
\end{abstract}
\pacs{  {\tt$\backslash$\string pacs\{13.25.Hw, 1.39.Fe, 14.40.Lb,
14.40.Nd \}} }
\newpage
\def\slash#1{#1 \hskip -0.5em / }

\section{Introduction}

The decays of heavy hadrons have recently received much attention in
the literature
\cite{IW1,BG,HG1,FGGW,FG,MRR1,ML,FGL,MRR2,IW2,HG2,stone,MN}. From the
experimental
standpoint, these decays allow access to some of the fundamental
parameters of the standard model, such as the elements of the
Cabibbo-Kobayashi-Maskawa (CKM) matrix. Questions of CP violation,
heavy-flavor oscillations and many others have added to this
interest.

{}From the theoretical standpoint, these processes, and the heavy
hadrons themselves, allow various quark models of QCD, as well
effective theories, to be tested. In particular, the heavy quark
effective theory (HQET) has both been tested by experimental
observations, and has played a major role in the extraction of $
\left|V_{cb}\right|$ from experimental data \cite{MN}.

Much of the success of HQET has been in the treatment of decays from
one heavy flavor to another, namely $b\to c$ transitions. The
effective theory is more limited in scope when applied to
heavy-to-light transitions, such as $c\to s$ or $b\to u$.
Neverthless, as we
will outline in a later section, the scaling behavior of the form
factors that describe various weak decays can be deduced \cite{iw2}.
This,
in principle, allows the form factors for $b\to s$ transitions to be
inferred from those for $c\to s$.

In this article, we assume the validity of the heavy quark symmetry
and examine the decays $D\to K^{(*)}\ell\nu$,
$B\to K^{(*)}\psi^{(\prime)}$, and the recently measured $B\to K^*
\gamma$ in a number of scenarios. In particular, we seek a scenario
in which all of these decays are adequately described by a single set
of form factors.
A number of authors have performed similar analyses
\cite{gourdin1,gourdin2,leyaouanc,carlson,cheng}, using the decays $D
\to K^{(*)}\ell\nu$ and
$B\to K^{(*)}\psi^{(\prime)}$, with varying results.

Once we find a scenario that is satisfactory for the decays mentioned
above, we examine, briefly, the decays $B\to K^{(*)}\ell^+\ell^-$.
In this way, we hope to make reliable estimates of the absolute decay
rates for
these processes. In a later article, we will consider more details of
these decays, such as forward-backward asymmetries. These decays,
as well as the decay $B\to K^*\gamma$, are particularly interesting
as the short distance operators responsible arise first at the
one-loop-level, and are therefore sensitive to new physics beyond
the standard model
\cite{raregroup1,raregroupa,raregroupb,raregroup2}. However, in order
for any effects due to such new physics to be clearly identified,
the long distance contributions
that arise in the hadronic matrix elements must be well understood.
As witness to this, we point out that the inclusive rate
$b\to s\gamma$ is reasonably well understood, while the exclusive rate
$B\to K^*\gamma$ has been predicted to be anywhere from 2\% to 40\%
of this inclusive rate \cite{playfer}.

In the case of the nonleptonic and the rare decays, there arises the
crucial issue of the form factors describing the $B\to K^{(*)}$
transition.
Such form factors may be estimated in various models, from QCD sum
rules, or by applying the scaling
relations predicted by HQET to the form factors for the corresponding
$D\to K^{(*)}$ transition. The method suggested by HQET, while
clearly
model-independent, is itself somewhat problematic, as one must take
into account the fact that predicting $B\to K^{(*)}\psi$, for instance,
requires that these relations be carefully extrapolated beyond the
kinematic range accessible in the $D\to K^{(*)}$
transitions. This is because the maximum $q^2$ in the $D\to K^{(*)}$
transitions is
1.95 (0.95) GeV$^2$, while the $q^2$ appropriate to the $B\to K^{(*)}
\psi$ transition is 9.6 GeV$^2$.

The HQET symmetry predictions relate form factors at the same value of
the kinematic variable $v\cdot v^\prime$ (or $v\cdot p^\prime$),
not $q^2$, where $v$ and
$v^\prime$ are the four-velocities of the parent and daughter hadron,
respectively. In this case,
the extrapolation is from $v\cdot v^\prime$ = 2.1 in $D\to K\ell\nu$
to $v\cdot v^\prime$ = 3.6 in $B\to K\psi$. The corresponding numbers
for the $K^*$ decays are 1.3 and 2.0.
The question of extrapolation also applies to the rare decays, and is
particularly important for the decay $B\to K^*\gamma$, for which
$q^2$ = 0,
but $v\cdot v^\prime$ = 3.0.

For the nonleptonic decays, a second issue is that of the
factorization approximation, which is commonly used to calculate the
hadronic matrix elements required. This approximation
is not very well founded in general theoretically,
yet it appears to work well phenomenologically in the $B$ decays where it
has been tested.
Nevertheless, application of the form factors of some model or
effective theory to the decay $B\to K\psi$, in conjunction with the
factorization approximation, serves to probe both issues, and may
fail due either to inadequate choice of form factors, failure of the
factorization approximation, or both.

The rest of this article is organized as follows. In the next section
we describe briefly the effective Hamiltonians and the
appropriate hadronic matrix elements for the processes of interest.
In section III we use HQET to obtain relations among the form factors
of
interest. In section IV we discuss our fitting procedure, and present
the results for the decays
that we are considering. In section V we discuss possible limitations
of our results, and suggest questions that may be of
interest to both theorists and experimentalists.

\section{Decay Processes}

\subsection{Semileptonic Decays}
\label{sldecays}

Of the three processes we discuss, the semileptonic decays are
perhaps the simplest to treat theoretically. The effective
Hamiltonian
for these decays is
\begin{equation}
{\cal H}_{\rm eff}=\frac{G_F}{\sqrt{2}}V_{cs}\bar{s} \gamma_\mu\left(1-
\gamma_5\right)c
\bar{\ell}\gamma^\mu\left(1-\gamma_5\right)\nu_\ell.
\end{equation}
The hadronic matrix elements for the decays $D\to K^{(*)}\ell\nu$ are

\begin{eqnarray}
\left<K(p^\prime)\left|\bar s\gamma_\mu c\right|D(p)\right>&=&f_+(p+p^
\prime)_\mu+f_-(p-p^\prime)_\mu,\nonumber\\
\left<K(p^\prime)\left|\bar s\gamma_\mu \gamma_5c\right|D(p)
\right>&=&0,\nonumber\\
\left<K^*(p^\prime,\epsilon)\left|\bar s\gamma_\mu c\right|D(p)
\right>&=&ig\epsilon_{\mu\nu\alpha\beta}
\epsilon^{*\nu}(p+p^\prime)^\alpha(p-p^\prime)^\beta,\nonumber\\
\left<K^*(p^\prime,\epsilon)\left|\bar s\gamma_\mu\gamma_5 c
\right|D(p)\right>&=&f\epsilon^*_\mu+a_+\epsilon^*\cdot p(p+p^
\prime)_\mu
+a_-\epsilon^*\cdot p(p-p^\prime)_\mu.
\end{eqnarray}
These decays are thus described in terms of six independent, {\it a
priori} unknown form factors. The terms in $f_-$ and $a_-$
are unimportant when the lepton mass is ignored, since
\begin{equation}
(p-p^\prime)_\mu\bar\ell\gamma^\mu(1-\gamma_5)\nu_\ell =(k_\nu+k_
\ell)_\mu\bar\ell\gamma^\mu(1-\gamma_5)\nu_\ell
=m_\ell\bar\ell\gamma^\mu(1-\gamma_5)\nu_\ell.
\end{equation}

In experimental analyses, the form factors $f_+$, $a_+$, $f$ and $g$ are
usually assumed to have the form
\begin{equation}
f_i(q^2)=\frac{f_i(0)}{1-\frac{q^2}{m_i^2}},
\end{equation}
where $q^2=(p-p^\prime)^2$, and $m_i$ is a mass, usually taken to be
that of the nearest resonance with the appropriate quantum numbers.
To date, only the mass appropriate to $f_+$ has been measured, and
its value measured by CLEO is $2.0\pm 0.22$ GeV \cite{cleo1}.
The masses appropriate to $a_+$, $f$ and $g$ are assumed to be 2.5
GeV, 2.5 GeV and 2.1 GeV respectively. The $f_i(0)$ have the
values $0.66\pm 0.03$, $-0.14\pm 0.03$, $0.40\pm 0.07$ and $1.55\pm
0.11$, for $f_+$, $a_+$, $g$ and $f$, respectively \cite{pdg}.

\subsection{Nonleptonic Decays}

Neglecting penguin contributions, the effective Hamiltonian for the
nonleptonic decays of interest here is
\begin{eqnarray}
{\cal H}_{\rm eff} &=& \frac{G_F}{\sqrt{2}} V_{cb} V_{cs}^* \left[
          C_1 (m_b)
          \left( \bar{c} \gamma_\mu (1 - \gamma_5) b \right)
          \left( \bar{s} \gamma^\mu (1 - \gamma_5) c \right) \right.
\nonumber
\\      &+& \left. C_2 (m_b)
          \left( \bar{s} \gamma_\mu (1 - \gamma_5) b \right)
          \left( \bar{c} \gamma^\mu (1 - \gamma_5) c \right)
          \right],
\end{eqnarray}
where
\begin{eqnarray}
C_1 (m_b) &=& \frac{1}{2} \left[
\left( \frac{ \alpha_s (m_b) }{ \alpha_s (m_W) } \right)^{-6/23} +
\left( \frac{ \alpha_s (m_b) }{ \alpha_s (m_W) } \right)^{12/23}
\right],\nonumber\\
C_2 (m_b) &=& \frac{1}{2} \left[
\left( \frac{ \alpha_s (m_b) }{ \alpha_s (m_W) } \right)^{-6/23} -
\left( \frac{ \alpha_s (m_b) }{ \alpha_s (m_W) } \right)^{12/23}
\right].
\end{eqnarray}
Here, $m_W$ is the mass of the $W$ boson, and $\alpha_s (\mu)$ is the
running coupling of QCD. This effective Hamiltonian mediates two
classes of $B$ nonleptonic decays. The first
class contains a $D$ in the final state: $B \to D X$ where $X$ may be
$D_s, D_s^*$.
The second class contains a light meson in the final state:
$B \to$ `$K$' $Y$ where $Y$ is now a charmonium state.

To evaluate the matrix elements of the effective Hamiltonian we
employ the factorization assumption. By Fierz rearrangement we
rewrite the effective Hamiltonian in
a form which is suitable for use with this assumption.
Both terms of the effective Hamiltonian contribute, in general, but
for the decays in which we are interested, only the
second term is of interest, and it may be written
\begin{eqnarray}
{\cal H}_{\rm eff} &=& \frac{G_F}{\sqrt{2}} V_{cb} V_{cs}^*
          \left( C_2 (m_b) + \frac{1}{N_c} C_1 (m_b) \right)\nonumber
\\
&&\times  \left( \bar{s} \gamma_\mu (1 - \gamma_5) b \right)
          \left( \bar{c} \gamma^\mu (1 - \gamma_5) c \right),
\end{eqnarray}
where $N_c$ is the number of colors.

At this point, it has become customary to replace the coupling
coefficient by a phenomenological constant $a_2$, whose
absolute value is measured to be approximately 0.24. The sign of $a_2$ is
not important for our
discussion at this point. It will become important if long distance
contributions to the dileptonic rare decays are included.

For the decay $B\to K^{(*)} J/\psi$, we therefore write, after using
factorization,
\begin{eqnarray} \label{mheff}
\left< J/\psi K^{(*)} \left|\vphantom{K^{(*)}}
 {\cal H}_{\rm eff} \right| B \right> &=&
          \frac{G_F}{\sqrt{2}} V_{cb} V_{cs}^*a_2 \left< J/\psi
\left| \bar{c} \gamma_\mu (1 - \gamma_5) c \right| 0 \right>\nonumber
\\
&& \times   \left< K^{(*)} \left| \bar{s} \gamma^\mu (1 - \gamma_5) b
 \right| B \right>.
\end{eqnarray}
The hadronic matrix elements $\left< K^{(*)} \left|\bar{s} \gamma^\mu
(1 - \gamma_5) b \right| B \right>$ are analogs of those of the
previous
subsection, so that the form factors required for these matrix
elements may, in principle, be obtained from the `semileptonic'
decays of
$B$ mesons into kaons. Such decays do not take place in the
standard model, but we can invoke heavy quark symmetries to relate
the
form factors needed to those for the semileptonic decays of $D$
mesons to kaons.

The remaining matrix element is
\begin{equation}
\left< J/\psi (p_\psi,\varepsilon_\psi)\left|  \bar{c} \gamma^\mu (1
- - \gamma_5) c  \right| 0 \right>=f_\psi m_\psi\varepsilon_\psi^\mu.
\end{equation}
The decay constant $f_\psi$ can be obtained from the leptonic width
of the appropriate charmonium vector resonance as
\begin{equation}
f_\psi=\sqrt{\Gamma_{\psi\to\ell^+\ell^-}\frac{27m_\psi}{16\alpha^2
\pi}},
\end{equation}
where we have ignored lepton masses. In this way, we find $f_
\psi$=0.382 GeV, and $f_{\psi^\prime}$=0.302 GeV.

\subsection{Rare Decays}

In the standard model, the effective Hamiltonian for the decay $b\to
s\gamma$ is \cite{raregroup1,raregroupb,raregroup}
\begin{equation}
{\cal H}_{\rm eff}=-\frac{G_F}{\sqrt{2}}\frac{e}{8\pi^2}
V_{ts}^*V_{tb}C_7(m_b)\bar{s} \sigma_{\mu\nu}
\left[m_b\left(1+\gamma_5\right)+m_s\left(1-\gamma_5\right)\right]b
F^{\mu\nu},
\end{equation}
where $F^{\mu\nu}$ is the electromagnetic field strength tensor, and
the term in $m_s$ may be safely ignored.
For the decay $b\to s\ell^+\ell^-$, the corresponding effective
Hamiltonian is
\begin{eqnarray}
{\cal H}_{\rm eff}&=&\frac{G_F}{\sqrt{2}}\frac{\alpha}{4\pi}V_{ts}^*V_{tb}
\left[
2i\frac{m_b}{q^2} C_7(m_b)\bar{s}\sigma_{\mu\nu }q^\nu (1+\gamma_5)b
\bar \ell\gamma^\mu \ell\right.\nonumber\\
&+&\left.\vphantom{\frac{4 G_F}{\sqrt{2}}} C_9(m_b)\bar{s}\gamma_\mu
\left(1-\gamma_5\right)b\bar{\ell}\gamma^\mu\ell
+C_{10}(m_b)\bar{s}\gamma_\mu\left(1-\gamma_5\right)b\bar{\ell}
\gamma^\mu\gamma_5\ell\right].
\end{eqnarray}
The Wilson coefficients $C_i$ are as in the article by Buras {\it et
al.} \cite{raregroupb}. For the discussion at hand, we have ignored
contributions that
arise from closed $q\bar q$ loops, although these may be easily
included.

The only new hadronic matrix elements that arise in the rare decays
of interest are
\begin{eqnarray}
\left<K(p^\prime)\left|\bar s\sigma_{\mu\nu} b\right|B(p)\right>&=&is
\left[\left(p+p^\prime\right)_\mu\left(p-p^\prime\right)_\nu-
\left(p+p^\prime\right)_\nu\left(p-p^\prime\right)_\mu\right],
\nonumber\\
\left<K^*(p^\prime,\epsilon)\left|\bar s\sigma_{\mu\nu} b\right|B(p)
\right>&=&\epsilon_{\mu\nu\alpha\beta}\left[g_+\epsilon^{*\alpha}
\left(p+p^\prime\right)^\beta
+g_-\epsilon^{*\alpha}\left(p-p^\prime\right)^\beta\right.\nonumber\\
&&\left.+h\epsilon^*\cdot p\left(p+p^\prime\right)^\alpha\left(p-p^
\prime\right)^\beta\right],
\end{eqnarray}
introducing four new form factors. Due to the relation
\begin{equation}
\sigma^{\mu\nu}\gamma_5=\frac{i}{2}\varepsilon^{\mu\nu\alpha\beta}
\sigma_{\alpha\beta},
\end{equation}
we can easily relate the matrix elements above to those in which the
current is $\bar s\sigma_{\mu\nu}\gamma_5b$. Experimentally, nothing
is known about
the form factors $s$, $g_\pm$ nor $h$.

\section{HQET And Form Factors}

Using the Dirac matrix representation of heavy mesons, one may treat
heavy-to-light transitions
using the same trace formalism that has been applied to
heavy-to-heavy transitions \cite{hqet,falk}. In the effective theory,
a $D$ meson
traveling with velocity $v$ is represented as \cite{hqet,falk}
\begin{equation}
D(v)\to\frac{1+\slash{v}}{2}\gamma_5\equiv M_D(v),
\end{equation}
with an identical representation for a $B$ meson. The meson states of
the effective theory are normalized so that
\begin{equation}
\left<{\cal D}(v^\prime)\left|\right.{\cal D}(v)\right>=2v_0\delta^3
\left({\bf p-p^\prime}\right).
\end{equation}
The states of QCD and HQET are therefore related by
\begin{equation}
\left|D(v)\right>=\sqrt{m_D}\left|{\cal D}(v)\right>.
\end{equation}
In all that follows, we will represent the states of full QCD as $
\left|D(v)\right>$, and the states of HQET as $\left|{\cal D}(v)
\right>$.

Let us consider transitions between such a heavy meson ($D$
meson) and a light meson (kaon) of spin $J$, through a generic flavor
changing current. In the effective theory, the matrix element of
interest is
\begin{equation}
\left<K^{(J)}(p,\eta)\left|\bar s\Gamma h_v^{(c)}\right|{\cal D}(v)
\right>={\rm Tr}\left\{\Xi\Gamma M_D(v)\right\}.
\end{equation}
$\Gamma$ is an arbitrary combination of Dirac $\gamma$ matrices, and
$\eta$ is the fully symmetric, traceless, transverse, $J$-index
tensor that represents the polarization of the state with spin $J$.
The matrix $\Xi$ must be the most general that can be constructed
from the
kinematic variables available, and Dirac $\gamma$ matrices. The most
general form for this is \cite{MR}
\begin{eqnarray}\label{formsgeneral}
\Xi&=&\eta^{\mu_1\dots\mu_J}v_{\mu_1}\dots v_{\mu_{J-1}}\nonumber\\
&\times&\left[v_{\mu_J}\left(\xi_1^{(J)}(v\cdot p)+\slash{p}
\xi_2^{(J)}(v\cdot p)\right)
+\gamma_{\mu_J}\left(\xi_3^{(J)}(v\cdot p)+\slash{p}\xi_4^{(J)}(v
\cdot p)\right)\right]\left(\matrix{1\cr\gamma_5}\right).
\end{eqnarray}
The $\xi_i$'s are uncalculable, nonperturbative functions of the
kinematic variable $v\cdot p$, and the 1 ($\gamma_5$) is present if
the resonance
$K^{(J)}$ has natural (unnatural) parity.

{}From this point on, let us limit the discussion to only two of the
kaon resonances, namely the ground-state pseudoscalar kaon itself,
and its
vector counterpart, $K^*$. The above then leads to
\begin{eqnarray} \label{ff1}
\left<K(p)\left|\bar s\Gamma h_v^{(c)}\right|{\cal D}(v)\right>&=&{
\rm Tr}\left\{\left(\xi_1+\slash{p}\xi_2\right)\gamma_5\Gamma
M_D(v)\right\},\nonumber\\
\left<K^*(p,\epsilon)\left|\bar s\Gamma h_v^{(c)}\right|{\cal D}(v)
\right>&=&{\rm Tr}\left\{\left[\left(\xi_3+\slash{p}\xi_4\right)
\epsilon^*\cdot v+\slash{\epsilon}^*\left(\xi_5+\slash{p}\xi_6\right)
\right]\Gamma
M_D(v)\right\}.
\end{eqnarray}

At this point, let us emphasize that the form factors $\xi_i$ are
independent of the form of $\Gamma$, and so are valid for decays
through the
left handed current ($\Gamma=\gamma_\mu(1-\gamma_5)$), as well as for
rare decays ($\Gamma=\sigma_{\mu\nu}(1\pm\gamma_5)$). These form
factors are also
independent of the mass of the heavy quark, and are therefore
universal functions. Thus, they are valid for $D\to K$ decays, as
well as for $B\to K$ decays.
This independence of the quark mass allows us to deduce, in a
relatively straightforward manner, the scaling behavior of the usual
form factors that
describe these transitions \cite{iw2}. We illustrate this by
examining one matrix element in detail.

Consider
\begin{eqnarray}\label{formsscaling}
&&\left<K(p)\left|\bar s\gamma_\mu c\right|D(p_0)\right>=f_+(p_0+p)_
\mu+f_-(p_0-p)_\mu\nonumber\\
&=&\sqrt{m_D}\left<K(p)\left|\bar s\gamma_\mu h_v^{(c)}\right|{\cal
D}(v)\right>=\sqrt{m_D}{\rm Tr}\left\{\left(\xi_1+\slash{p}\xi_2
\right)\gamma_5\gamma_\mu
\frac{1+\slash{v}}{2}\gamma_5\right\},\nonumber\\
&=&2\sqrt{m_D}\left(\xi_1 v_\mu-\xi_2 p_\mu\right).
\end{eqnarray}
{}From these equations, one finds that
\begin{eqnarray}
\xi_1&=&\frac{\sqrt{m_D}}{2}\left(f_++f_-\right),\nonumber\\
\xi_2&=&\frac{1}{2\sqrt{m_D}}\left(f_--f_+\right).
\end{eqnarray}
Since the $\xi_i$ do not scale with the mass of the heavy quark (or
meson), it is trivial to deduce that
\begin{equation}
f_++f_- \approx m_D^{-1/2},\,\,\,\, f_+-f_- \approx m_D^{1/2}.
\end{equation}

For the other transitions of interest, the form factors are as
defined in the previous section, and the relationships between these
and the
$\xi_i$ are
\begin{eqnarray}\label{xitof}
\xi_1&=&\frac{\sqrt{m_D}}{2}\left(f_++f_-\right),\nonumber\\
\xi_2&=&\frac{1}{2\sqrt{m_D}}\left(f_--f_+\right)=-\sqrt{m_D} s,
\nonumber\\
\xi_3&=&\frac{m_D^{3/2}}{2}\left(a_++a_-\right),\nonumber\\
\xi_4&=&\frac{\sqrt{m_D}}{2}\left(2g-a_++a_-\right)=m_D^{3/2}h,
\nonumber\\
\xi_5&=&-\frac{1}{2\sqrt{m_D}}\left(f+2m_Dv\cdot p g\right)=-\frac{
\sqrt{m_D}}{2}\left(g_++g_-\right),\nonumber\\
\xi_6&=&\sqrt{m_D}g=\frac{1}{2\sqrt{m_D}}\left(g_--g_+\right).
\end{eqnarray}
These expressions yield the additional scaling behavior
\begin{eqnarray}
&&a_++a_-\approx m_D^{-3/2},\,\,\, a_+-a_-\approx m_D^{-1/2},
\nonumber\\
&&f\approx m_D^{1/2},\,\,\, g\approx m_D^{-1/2},\nonumber\\
&&s\approx m_D^{-1/2},\,\,\, h\approx m_D^{-3/2},\nonumber\\
&&g_++g_-\approx m_D^{-1/2},\,\,\,g_+-g_-\approx m_D^{1/2}.\nonumber
\end{eqnarray}

Eqns. (\ref{formsgeneral}) and (\ref{ff1}), and consequently eqns.
(\ref{formsscaling}-\ref{xitof}), contain all of the leading order
$m_D$ dependence in the form factors, and are valid irrespective of
the mass of the strange quark. If the strange quark could be treated as
heavy, then we could think of the $\xi_i$ as arising from an infinite
sum of terms in the $1/m_s$ expansion. The leading order forms (in
$1/m_s$) are also contained in these expressions. These expressions
are also valid in the limit of a very light strange quark.

Isgur and Wise \cite{iw2} have used the scaling of $f_++f_-$ ($\approx
m_D^{-1/2}$) to say that this combination of form factors
vanishes (at leading order in $1/m_D$), and suggest that one can write
$f_-=-f_++{\cal O}(1/m_D)$. Implicit in this argument is the assumption
that the strange quark is very light. In our formalism, this
amounts to setting $\xi_1$ to zero, and we would automatically lose
the full scope of our predictions. This is because we could then
never recover the limit of a heavy strange quark, for in this limit,
$\xi_1=-\sqrt{m_K}\xi/2$, where $\xi$ is the usual Isgur-Wise function.

The strange quark is such that it may be treated as either heavy or light.
We believe that neither the full heavy $s$ limit
($\xi_1=-\sqrt{m_K}\xi/2$), nor
the limit of a very light $s$ quark ($\xi_1\to 0$) is completely
satisfactory. We assume neither limit in our analysis, and therefore
make full use of the predictions of the heavy quark effective theory,
which are valid independent of the mass of the strange quark. This means
that we retain the form factor $\xi_1$ in our discussion and treat it as
a completely independent form factor, tying it to neither of the two
limits discussed.

At the risk of boring the overly patient reader to tears, and perhaps
even to death, we list one more set of relationships among form
factors, this
time writing the usual form factors in terms of the $\xi_i$. The
relations are
\begin{eqnarray} \label{ff2}
f_+&=&\frac{1}{\sqrt{m_D}}\left(\xi_1-m_D\xi_2\right),\nonumber\\
f_-&=&\frac{1}{\sqrt{m_D}}\left(\xi_1+m_D\xi_2\right),\nonumber\\
f&=&2\sqrt{m_D}\left(\xi_5+v\cdot p\xi_6\right),\nonumber\\
a_+&=&-\frac{1}{m_D^{3/2}}\left[\xi_3+m_D\left(\xi_6-\xi_4\right)
\right],\nonumber\\
a_-&=&-\frac{1}{m_D^{3/2}}\left[\xi_3-m_D\left(\xi_6-\xi_4\right)
\right],\nonumber\\
g&=&\frac{1}{\sqrt{m_D}}\xi_6.
\end{eqnarray}
For the corresponding transitions with a $b$ quark (and a
$B$ meson) in the initial state, all factors of $m_D$ above must be
replaced by
$m_B$. Using this, rearrangement of eqns. (\ref{ff2}) yields
\begin{eqnarray} \label{ffmix}
f_+^B(v\cdot p)&=&\frac{1}{2}\left(\frac{m_B}{m_D}\right)^{1/2}
\left[f_+^D(v\cdot p)\left(1+\frac{m_D}{m_B}\right)
+f_-^D(v\cdot p)\left(\frac{m_D}{m_B}-1\right)\right],
\nonumber\\
f_-^B(v\cdot p)&=&\frac{1}{2}\left(\frac{m_B}{m_D}\right)^{1/2}
\left[f_-^D(v\cdot p)\left(1+\frac{m_D}{m_B}\right)
+f_+^D(v\cdot p)\left(\frac{m_D}{m_B}-1\right)\right],
\nonumber\\
f^B(v\cdot p)&=&\left(\frac{m_B}{m_D}\right)^{1/2}f^D(v\cdot p),
\nonumber\\
g^B(v\cdot p)&=&\left(\frac{m_D}{m_B}\right)^{1/2}g^D(v\cdot p),
\nonumber\\
a_+^B(v\cdot p)&=&\frac{1}{2}\left(\frac{m_D}{m_B}\right)^{1/2}
\left[a_+^D(v\cdot p)\left(1+\frac{m_D}{m_B}\right)
+a_-^D(v\cdot p)\left(\frac{m_D}{m_B}-1\right)\right],
\nonumber\\
a_-^B(v\cdot p)&=&\frac{1}{2}\left(\frac{m_D}{m_B}\right)^{1/2}
\left[a_-^D(v\cdot p)\left(1+\frac{m_D}{m_B}\right)
+a_+^D(v\cdot p)\left(\frac{m_D}{m_B}-1\right)\right],
\end{eqnarray}
where $f_+^D$ is the form factor appropriate to the $D\to K$
transition, while $f_+^B$ is the form factor appropriate to the $B\to
K$
transition, and quantities on the left-hand-sides of eqns.
(\ref{ffmix}) are evaluated at the same values of $v\cdot p$ as those
on
the right-hand-sides. Omitted from each of eqn. (\ref{ffmix}) is a QCD
scaling factor, discussed below.

Eqns. (\ref{ffmix}) illustrate two effects, namely the scaling of form
factors in going from the $D$ system to the $B$ system, as well as
the
mixing of $a_+$ with $a_-$, and $f_+$ with $f_-$. The effect of this
mixing is very important in going from $D$ transitions to $B$
transitions,
as it introduces form factors that have not yet been measured
experimentally, or to which experiments are not yet sensitive,
namely $f_-$ and $a_-$. In the rates for the semileptonic decays $D
\to K^{(*)}\ell\nu$, terms dependent on $f_-$ and $a_-$ are
proportional to the mass of the lepton, and
thus play a miniscule role, except near $q^2=m_\ell^2$. Such terms
may also be significant if the polarization of the charged lepton is
measured.

We close this section with a brief discussion of radiative
corrections to the currents and matrix elements that we have
discussed. In the
limit of a heavy $b$ quark, the full current of QCD is replaced by
\cite{qcd}
\begin{equation}
\bar s\Gamma b\to \bar s \Gamma h_v^{(b)}\left[\frac{\alpha_s(m_b)}{
\alpha_s(\mu)}\right]^{-\frac{6}{25}}.
\end{equation}
This arises from integrating out the $b$ quark,
and matching the resulting effective theory onto full QCD at the
scale $m_b$, at one loop level. At the scale $m_c$, we must also
integrate out the $c$ quark, but there is also the effect due to
running between $m_b$ and $m_c$. The net effect of this is that
the form factors $\xi_i$ appropriate to the $b\to s$
transitions are related to those for the $c\to s$ transitions by
\begin{equation}
\xi_i^{b\to s}=\xi_i^{c\to s}\left[\frac{\alpha_s(m_b)}{
\alpha_s(m_c)}\right]^{-\frac{6}{25}}.
\end{equation}

The forms of the matrix elements that we
discuss above are valid in the limit of infinitely heavy $b$ and $c$
quarks. For quarks of finite mass, there are clearly going to be
corrections to the relations we have obtained. In other words,
new form factors that appear first at order $1/m_c$ and $1/m_b$ will
begin to make contributions. It is expected that such contributions
will become more significant away from the `non-recoil' point, or for
$v\cdot p >m_{K^{(*)}}$. This is particularly important for the
$B\to K^{(*)}$ decays, as $v\cdot p$ can become very large.
Nevertheless, we will apply the relations we have found through all
of the available phase space. It is our hope
that by doing this, we will at least be able to indicate the
suitability of HQET for these transitions. However, since
we
fit the form factors rather than attempt to calculate them in some
model, some of these higher order effects may, in fact, have been
included.

\section{Results And Discussion}

\subsection{Data}

All of the results we describe are for decays with $\bar K^0$ (or
$\bar K^{0*}$) in the final state. The treatment of charged kaons
would be identical, and we believe that our results in these channels
would be similar
in quality to those we obtain for neutral kaons.  Before describing
the results of our fits, we must comment on how we treat the
available data, particularly in the case of the semileptonic
decays. Very few of the experimental collaborations have extracted
acceptance-corrected distributions \cite{cleo1,exper}. Instead, the
form factors are all assumed to
be of the monopole form \cite{pdg}, and the parameters are then
extracted from the Monte Carlo simulations, with acceptances and
efficiencies folded in.

Because of this, we proceed in the inverse sense to generate some
`simulated data'. We use the published monopole
parameters for the form factors to generate $d\Gamma/dq^2$ spectra
for the semileptonic decays, using the published uncertainties in the
monopole parameters to generate uncertainties in the simulated data.
In general, these
errors are correlated, but we ignore this correlation.

The simulated data generated in this way are completely smooth. We
introduce an `anti-smoothing' by smearing the simulated data  with a
pseudo-randomly generated gaussian distribution of mean zero and
standard deviation determined by the errors in the unsmeared
simulated data. It is this smeared simulated data that we use for
fitting. For the decays $D^+\to\bar K^{0*}\ell^+\nu$, we also include
the ratios $\Gamma_L/\Gamma$ and $\Gamma_+/\Gamma_-$ in the fit,
where $\Gamma_\pm$ are as defined in PDG \cite{pdg}.
In addition, we must point out that the measured decay rates for
$D^+\to \bar K^{0(*)}\ell^+\nu$ are somewhat smaller than those for
$D^0\to \bar K^{-(*)}\ell^+\nu$, while the published form factor
parametrizations
are for the average over charge states. To account for this, we
rescale the values of the $f_i(0)$ to correspond to the smaller rates
for neutral kaons.
It is these rescaled values that are cited in section \ref{sldecays},
and that we use to generate the simulated data.

For the nonleptonic decays, we fit to the PDG averaged decay widths
for $\bar B^0\to \bar K^0\psi$ and $\bar B^0\to \bar K^{0*}\psi$. In
the case of the
latter, we also include the ratio $\Gamma_L/\Gamma$. For this ratio,
we take the averaged value of $0.78\pm 0.073$ as calculated by
Gourdin {\it et al.} \cite{gourdin1,gourdin2}.
Masses and lifetimes of mesons are all taken from PDG \cite{pdg}, and
we use $V_{tb}=0.9988$, $V_{ts}=0.03$, $V_{cs}=0.9738$,
$V_{cb}=0.041$, $m_b=4.9$ GeV,
$m_c$=1.5 GeV, $m_t$=177 GeV.

It is worth mentioning that the experimental choice of monopole form
factors may be inappropriate, particularly for $f$.
In the limit of a heavy strange quark, one finds that $f\propto
\left(1+v\cdot v^\prime\right)
\xi(v\cdot v^\prime)$, where $\xi$ is the Isgur-Wise function. If $
\xi$ is assumed to be monopolar in $q^2$, then simple
pole dependence for $f$ is inappropriate. Even in the case of a light
strange quark, we find that $f\propto \xi_5+
v\cdot p \xi_6$, again indicating that the dependence on the
kinematic variable is not simply a monopole form.
However, for the range of $q^2$ accessible in $D\to
K^*$ decays, the decay rate is not sensitive to the
kinematic dependence. The effect on the decay rate for the
nonleptonic and rare decays being considered here, however, are quite
significant.

In our fitting, we have separated the decays
containing a $K$ meson in the final state from those containing a
$K^*$ meson in the final state.
Thus, for instance, we do not include data on ratios of rates like $
\Gamma(D\to K^*\ell\nu)/\Gamma(D\to K\ell\nu)$. Our reason is that
such ratios
introduce correlations between the parameters of the form factors for
the two sets of decays.

\subsection{Heavy-$s$ Limit}

One approximation used recently in the literature has been to treat
the strange quark as heavy \cite{liu,mannel}, so that the decays of
interest can be treated
in the heavy-to-heavy limit. In this limit, the form factors of eqn.
(\ref{ff1}) may be written
\begin{eqnarray}
\xi_1&=&\xi_5=\frac{\sqrt{m_K}}{2}\xi,\nonumber\\
- -\xi_2&=&\xi_6=\frac{1}{2\sqrt{m_K}}\xi,\nonumber\\
\xi_3&=&\xi_4=0,
\end{eqnarray}
where $\xi$ is the Isgur-Wise function for heavy-to heavy
transitions, and in this limit, $m_K=m_{K^*}$. In particular, this
limit means that $\xi_5=m_K\xi_6$.

We have used this form in fitting the data, and have obtained
reasonable fits to the differential decay rates in the semileptonic
decays, as well
as to the total decay rates in the nonleptonic decays. Polarization
ratios, however, are poorly reproduced. In the case of the
nonleptonic
decay $B\to K^*\psi$, the ratio $\Gamma_L/\Gamma$ depends only on
kinematics, and has a value of 0.43, independent of the form chosen
for
$\xi$. The experimental value is $0.78\pm 0.07$. In addition, the
ratio $\Gamma_+/\Gamma_-$ in $D\to K^*\ell\nu$ always has a value of
about 0.4, significantly different from the experimental value
of 0.16$\pm$0.04, and largely independent of the form chosen for $
\xi$. This indicates that the value of 0.43 obtained for the ratio
$\Gamma_L/\Gamma$ in $B\to K^*\psi$ is not necessarily due to the
breakdown of factorization in the heavy $s$ limit, as this limit
does not even provide an adequate description of all measurements in
the semileptonic decays.

Relaxing the strict heavy-$s$ limit, but constraining the form
factors to be near this limit, does not help much, as the
polarization observables
are still not well reproduced. The conclusion that we draw from this is
that the heavy-$s$ limit may give an acceptable description of
unpolarized data,
but may be dangerous when applied to polarization observables.

\subsection{General Features Of Results}

All of the results we present are obtained by performing four kinds
of fits, namely (1) include the semileptonic decays $D\to K^{(*)}\ell
\nu$
only; (1) include the semileptonic decays as well as the nonleptonic
decays $B\to K^{(*)}\psi$; (3) include the semileptonic decays, the
nonleptonic
decays $B\to K^{(*)}\psi$, and the nonleptonic decays $B\to K^{(*)}
\psi^\prime$; (4) include the measured decay rate for $B\to K^*
\gamma$ in the fit, as well as
the three other decays.
Clearly, in the case of the decays to $K$ mesons, we need perform
only  three kinds of fits. In cases where a measured quantity is not
included in the
fit, we calculate that quantity using the fit parameters.

We have explored two sets of parametrizations of the form factors. In
the first scenario, which we refer to as the exponential scenario,
each $\xi_i$ has one of the forms
\begin{eqnarray}
\xi_i&=&a_i\exp{\left[-b_i\left(v\cdot p-m_{K^{(*)}}\right)
\right]}=a_i\exp{\left[-\frac{b_i}{2m_D}\left(q^2_{{\rm max}}
- -q^2\right)\right]}\label{exp1},\\
\xi_i&=&a_i\exp{\left[-b_i\left(v\cdot p-m_{K^{(*)}}\right)^2
\right]}=a_i\exp{\left[-\frac{b_i}{4m_D^2}\left(q^2_{{\rm max}}-q^2
\right)^2\right]}
\label{exp2},\\
\xi_i&=&a_i\exp{\left[-b_i\left(v\cdot p\right)^2\right]}\label{exp3},
\end{eqnarray}
while in the second scenario, which we call the multipolar scenario,
the forms chosen are
\begin{equation}
\xi_i=a_i\left(1+b_iv\cdot p\right)^{n_i},
\end{equation}
with $n_i=-2,\,-1,\,0,\,1$. In the exponential scenario, eqn.
(\ref{exp1}) most closely corresponds to the forms that arise in some
quark models, most notably that of Isgur and collaborators \cite{isgw}.
However, in such models, the exponential of eqn. (\ref{exp1}) is
usually multiplied by a polynomial in $v\cdot p$, or equivalently,
in $q^2$. In the multipolar scenario, $n_i=-2$
corresponds to a dipole form factor, while $n_i=-1$ represents a
monopole.
In any one fit, we do not choose all the form factors to have the
same form. This means that, for instance, in the
case of the decays to $K$ mesons, for which there are two form
factors, the second scenario corresponds to sixteen different
possible combinations of
forms for $\xi_1$ and $\xi_2$.

In each case, the $a_i$ and $b_i$ are the free parameters to be
varied in the fit. There are therefore twelve free parameters in each
fit, to be compared with
five extracted and three assumed parameters in the measured
semileptonic decays. Since we have more free parameters than are
extracted from the experimental data,
one might expect that it should be very easy to account for all of
the data. In fact, the number of free parameters poses some problems,
as it means that the
problem is not very well constrained. One consequence of this is that
there appear to be several local minima for any particular choice of
form factors. Nevertheless,
there are some combinations of form factors that simply do not
provide adequate descriptions of the data, despite the large number
of free parameters.

When only the semileptonic decays are included in the fit, we find
that almost any combination of forms for the form factors leads to
reasonable results.
The few combinations that do not provide good descriptions fail only
in their description of the polarization observables.

When we use the exponential forms, we find that we are not able to
obtain an adequate description of all of the data simultaneously.
In particular, when we include the
nonleptonic decays in the fit, the prediction for the rare decay $B
\to K^*\gamma$ is significantly different from the measured rate.
In some cases, however, we find that if we fit the semileptonic
decays alone, omitting all of the nonleptonic decays, the prediction
for the rare decays is of the right order of magnitude. One possible
conclusion here is that factorization is not applicable to these
nonleptonic decays, or that there are significant non-factorizable
contributions
to the amplitude.

In contrast with the exponentials of the first scenario, many
combinations of the `multipolar' forms of the second scenario lead to
good descriptions of all the
data simultaneously. One outstanding feature of all
of our results in this scenario (which also exists in the exponential
scenario, but to a lesser extent) can be easily understood by
examining eqn. (\ref{ff2}),
where it is seen that $\xi_6$ is present in all of the form factors
that describe the semileptonic decays $D\to K^*$. It is therefore not
at all surprising that our results are most sensitive to this form
factor. Invariably, we have found that the best fits occur for $
\xi_6$ linear in $v\cdot p$ ({\it i.e.}, $n_6=1$), independent of the
forms chosen for the
other form factors. Furthermore, the slope parameter $b_6$ is almost
always negative, with values lying between -0.4 and -0.65 GeV$^{-1}$.
The only
positive values for $b_6$ occur when only the
semileptonic decays are included.

The fact that $\xi_6$ is so well constrained in our fits means that
$f$ and $g$ are also quite well constrained. Since these are the only
two form factors that are needed for the decay $B\to K^*\gamma$, it
is no surprise that we find little variation in the predictions for
this decay rate as we vary the parametrizations of $\xi_3$, $\xi_4$
and $\xi_5$, provided that the choice of parametrization of $\xi_6$
is unchanged.

For other forms of $\xi_6$, we find that at most two of the
semileptonic, nonleptonic or rare decays are well accomodated. For
instance, if we choose a monopole form for $\xi_6$, then in
addition to the semileptonic decays, we find that we can accomodate
either the nonleptonic decays, the rare decay, but not both. In addition,
if we choose all form
factors to be monopolar, we fail to find adequate descriptions of the
polarization observables, particularly for the ratio $\Gamma_L/
\Gamma$ measured in the
nonleptonic decay.

Our results for the form factors are comparable with those of other
authors. Gourdin {\it et al.} \cite{gourdin1} have found that a number of
scenarios, including monopolar form factors, are unable to describe the
nonleptonic measurements. Aleksan {\it et al.} \cite{leyaouanc} have
found that softening the scaling relations allows an adequate description,
while in a second article, Gourdin {\it et al.} \cite{gourdin2} have found
that allowing the form factor $f$ to decrease linearly with $q^2$ allows
an adequate description of the data. We find that $f$ is quadratic in
$q^2$, but the absolute value at $q^2=0$ for the $B\to K^*$ transitions
is larger than the absolute value at $q^2=q^2_{\rm max}$, in keeping
with the results of \cite{gourdin2}, and quite different from pole models.

\subsection{Parameters And Form Factors}
\squeezetable

For each scenario, we have selected a set of fits that we consider to
be representative. The parameters for these fits are displayed in
table
\ref{fitparame} for the exponential forms, and in table
\ref{fitparam} for the multipolar forms. In the first scenario, the
results we have selected
correspond to
$\xi_2$ and $\xi_3$ as in eqn. (\ref{exp1}), $\xi_1$ and $\xi_4$ as
in eqn. (\ref{exp2}), and  $\xi_5$ and $\xi_6$ as in eqn. (
\ref{exp3}). In the second
scenario, the $n_i$ have the values $n_i=(0,-1,1,-2,1,1)$.

\begin{table}
\caption{Values of the parameters that result from four different
fits, for the exponential scenario. In this table,
Fit 1 means that only $D\to K^{(*)}\ell\nu$ is included in the fit;
Fit 2 means $D\to K^{(*)}\ell\nu$ and $B\to K^{(*)}J/\psi$ are
included; Fit 3 means $D\to K^{(*)}\ell\nu$, $B\to K^{(*)}J/\psi$ and
$B\to K^{(*)}\psi^\prime$ are included; Fit 4
means $D\to K^*\ell\nu$, $B\to K^*J/\psi$, $B\to K^*\psi^\prime$ and
$B\to K^*\gamma$ are all included, and applies only to decays with
$K^*$'s in the final state.
\label{fitparame}}
\begin{tabular}{|c||c|c|c|c||}\hline
Parameter & Fit 1 & Fit 2 & Fit 3 & Fit 4\\ \hline
$a_1$ & 2.497 & 0.671 & 0.667 & - \\ \hline
$a_2$ & 1.503 & -0.265 & -0.266 & - \\ \hline
$a_3$ & 10.0 & 8.825 & 1.745 & 1.742 \\ \hline
$a_4$ & 4.521 & 5.137 & 1.259 & 1.257 \\ \hline
$a_5$ & 9.996 & 9.998 & 9.990 & 9.726 \\ \hline
$a_6$ & 1.044 & 1.037 & 1.080 & 1.074 \\ \hline
$b_1$ & 3.353 & 1.979 & 1.984 & - \\ \hline
$b_2$ & 4.710 & 8.0$\times 10^{-5}$ & 3.1$\times 10^{-6}$ & - \\
\hline
$b_3$ & 3.511 & 0.238 & 7.6$\times 10^{-5}$ & 3.0 $\times 10^{-6}$ \\
\hline
$b_4$ & 10.0 & 1.447 & 1.076 & 1.076 \\ \hline
$b_5$ & 5.914 & 6.730 & 6.520 & 6.458 \\ \hline
$b_6$ & 0.585 & 0.563 & 0.585 & 0.581 \\ \hline
\end{tabular}
\end{table}

\begin{table}
\caption{Values of the parameters that result from four different
fits, for the multipolar scenario. The columns are as in table
\protect\ref{fitparame}.}
\label{fitparam}
\begin{tabular}{|c||c|c|c|c||}\hline
Parameter & Fit 1 & Fit 2 & Fit 3 & Fit 4\\ \hline
$a_1$ & 3.358 & 0.144 & 0.158 & - \\ \hline
$a_2$ & 2.929 & 5.875 & 5.951 & - \\ \hline
$a_3$ & -10.0 & 0.162 & 0.144 & 0.145 \\ \hline
$a_4$ & -10.0 & 3.154 & 2.569 & 2.365 \\ \hline
$a_5$ & 1.399 & 0.288 & 0.346 & 0.451 \\ \hline
$a_6$ & 0.055 & 1.094 & 1.085 & 0.922 \\ \hline
$b_1$ & - & - & - & - \\ \hline
$b_2$ & 0.290 & 10.0 & 10.0 & - \\ \hline
$b_3$ & -0.507 & 10.0 & 10.0 & 10.0 \\ \hline
$b_4$ & 1.134 & 0.468 & 0.388 & 0.354 \\ \hline
$b_5$ & -1.012 & -1.042 & -1.030 & -0.907 \\ \hline
$b_6$ & 10.0 & -0.439 & -0.435 & -0.392 \\ \hline
\end{tabular}
\end{table}
To give some sort of meaning to these parameters, we display in
tables \ref{fitparame1} and \ref{fitparam1}, the values of the form
factors at $q^2=0$,
as well as their logarithmic derivatives at the same point. In these
tables, the theoretical error that we quote, as well as those that we
quote for
the rest of our results, are estimates only, and are obtained by
using the covariance matrix that results from the fit.

For the monopole form chosen by experimentalists, the logarithmic
derivative is
\begin{equation}
\left.\frac{1}{f_i(q^2)}\frac{df_i(q^2)}{dq^2}\right|_{q^2=0}=
\frac{1}{m_i^2},
\end{equation}
where $m_i$ is the polar mass. In this way, we can compare our form
factors at $q^2=0$, and the corresponding slope parameters, with the
experimental
values. It is gratifying to find that for $f$, $g$ and $f_+$, the
values we have obtained are, for the most part, quite consistent with
the
experimental values, for both the exponential and multipolar
scenarios. In addition, the logarithmic derivative of $a_+$
shows some variation as we go from fit to fit, and there are sizable
variations
in the values for $f_-$, $a_-$ and their logarithmic derivatives.
These variations are not surprising, as there are no experimental
constraints on these quantities. Although the values we obtain for
these quantities may lie
outside the accepted domain suggested by models, it is nevertheless
quite satisfying to note that the values do not change much as we go
from Fit 2 to
Fit 3 to Fit 4, particularly for the multipolar scenario. It is also
interesting that the prefered slope parameter for $a_+$ is
negative.

\begin{table}
\caption{Values of the form factors and their logarithmic derivatives
at $q^2=0$, for the exponential scenario. The columns are as in table
\protect\ref{fitparame}.}
\label{fitparame1}
\begin{tabular}{|c||c||c|c|c|c||}\hline
Quantity & Experiment & Fit 1 & Fit 2 & Fit 3 & Fit 4\\ \hline
$f_+(0)$ & 0.66$\pm$0.03 & 0.59$\pm$0.04 & 0.66$\pm$0.03 & 0.66$
\pm$0.03 & - \\ \hline
$\frac{f^\prime_+(0)}{f_+(0)}$ & $0.23\pm 0.09$ & 0.80$\pm$0.19 &
0.24$\pm$0.07 & 0.24$\pm$0.07 & -\\ \hline
$f_-(0)$ & - & 0.97$\pm$0.48 & -0.06$\pm$0.12 & -0.07$\pm$0.12 & - \\
\hline
$\frac{f^\prime_-(0)}{f_-(0)}$ & - & 0.98$\pm$0.17 & -2.50$\pm$4.61 &
- -2.32$\pm$3.96 & - \\ \hline
$f(0)$ & 1.55$\pm$0.11 & 1.53$\pm$0.10 & 1.55$\pm$0.07 & 1.57$
\pm$0.11 & 1.57$\pm$0.08 \\ \hline
$\frac{f^\prime(0)}{f(0)}$ & 0.16 & 0.15$\pm$0.12 &  0.12$\pm$0.05 &
0.14$\pm$0.03 & 0.14$\pm$0.05 \\ \hline
$g(0)$ & 0.40$\pm$0.07 & 0.35$\pm$0.03 & 0.36$\pm$0.02 & 0.37$
\pm$0.03 & 0.37$\pm$0.02 \\ \hline
$\frac{g^\prime(0)}{g(0)}$ & $0.23$ & 0.36$\pm$0.12 & 0.35$\pm$0.05 &
0.36$\pm$0.02 & 0.36$\pm$0.05 \\ \hline
$a_+(0)$ & -0.14$\pm$0.03 & -0.23$\pm$0.05 & -0.20$\pm$0.05 & -0.19$
\pm$0.05 & -0.19$\pm$0.04 \\ \hline
$\frac{a^\prime_+(0)}{a_+(0)}$ & 0.16 & -3.17$\pm$0.66 & -1.72$
\pm$0.76 & 0.03$\pm$0.11 & 0.03$\pm$0.44 \\ \hline
$a_-(0)$ & - & -2.99$\pm$0.36 & -6.33$\pm$1.98 & -1.18$\pm$0.05 &
- -1.18$\pm$0.28 \\ \hline
$\frac{a^\prime_-(0)}{a_-(0)}$ & - & 1.25$\pm$0.15 & 0.12$\pm$0.08 &
- -0.05$\pm$0.01 & -0.05$\pm$0.11 \\ \hline
\end{tabular}
\end{table}

\begin{table}
\caption{Values of the form factors and their logarithmic derivatives
at $q^2=0$, for the multipolar scenario. The columns are as in table
\protect\ref{fitparame}.}
\label{fitparam1}
\begin{tabular}{|c||c||c|c|c|c||}\hline
Quantity & Experiment & Fit 1 & Fit 2 & Fit 3 & Fit 4\\ \hline
$f_+(0)$ & 0.66$\pm$0.03 & 0.64$\pm$0.06 & 0.63$\pm$0.02 & 0.62$
\pm$0.04 & - \\ \hline
$\frac{f^\prime_+(0)}{f_+(0)}$ & $0.23\pm 0.09$ & 0.29$\pm$0.17 &
0.29$\pm$0.07 & 0.29$\pm$0.08 & -\\ \hline
$f_-(0)$ & - & -5.56$\pm$1.88 & -0.84$\pm$0.34 & -0.86$\pm$0.34 & -
\\ \hline
$\frac{f^\prime_-(0)}{f_-(0)}$ & - & 0.03$\pm$0.03 & 0.21$\pm$0.04 &
0.21$\pm$0.04 & - \\ \hline
$f(0)$ & 1.55$\pm$0.11 & 1.55$\pm$0.14 & 1.55$\pm$0.08 & 1.53$
\pm$0.18 & 1.54$\pm$0.07 \\ \hline
$\frac{f^\prime(0)}{f(0)}$ & 0.16 & 0.04$\pm$0.12 & 0.14$\pm$0.06 &
0.17$\pm$0.20 & 0.15$\pm$0.05 \\ \hline
$g(0)$ & 0.40$\pm$0.07 & 0.51$\pm$0.05 & 0.40$\pm$0.03 & 0.40$
\pm$0.05 & 0.37$\pm$0.03 \\ \hline
$\frac{g^\prime(0)}{g(0)}$ & $0.23$ & 0.22$\pm$0.01 & 0.24$\pm$0.03 &
0.23$\pm$0.05 & 0.19$\pm$0.02 \\ \hline
$a_+(0)$ & -0.14$\pm$0.03 & -0.24$\pm$0.05 & -0.21$\pm$0.04 & -0.20$
\pm$0.08 & -0.20$\pm$0.03 \\ \hline
$\frac{a^\prime_+(0)}{a_+(0)}$ & 0.16 & -1.14$\pm$0.33 & -1.11$
\pm$0.30 & -0.93$\pm$0.71 & -0.98$\pm$0.27 \\ \hline
$a_-(0)$ & - & 3.54$\pm$0.33 & -1.37$\pm$0.25 & -1.21$\pm$0.61 &
- -1.21$\pm$0.22 \\ \hline
$\frac{a^\prime_-(0)}{a_-(0)}$ & - & 0.22$\pm$0.02 & -0.08$\pm$0.01 &
- -0.09$\pm$0.11 & -0.09$\pm$0.01 \\ \hline
\end{tabular}
\end{table}

\begin{figure}
\centerline{\mbox{\begin{turn}{0}%
\epsfxsize=3.0in\epsffile{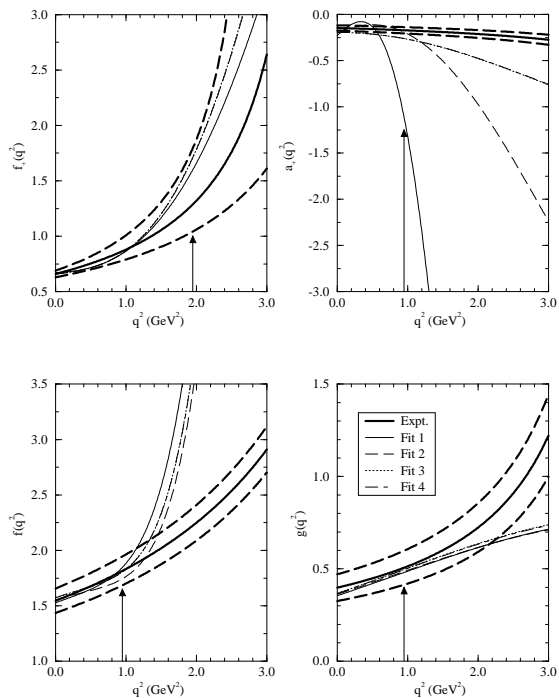}\end{turn}}}
\vskip 0.25in
\caption{Form factors for the $D\to K^{(*)}$ transitions that result
from fits, using exponential forms. The thick solid curve is the
experimentally extracted form
factor, while the thick dashed curves show the range that results
from the experimental uncertainties in the parametrizations of the
form factors. In each graph,
Fit 1 means that only $D\to K^{(*)}\ell\nu$ is included in the fit;
Fit 2 means $D\to K^{(*)}\ell\nu$ and $B\to K^{(*)}J/\psi$ are
included; Fit 3 means $D\to K^{(*)}\ell\nu$, $B\to K^{(*)}J/\psi$ and
$B\to K^{(*)}\psi^\prime$ are included; Fit 4
means $D\to K^*\ell\nu$, $B\to K^*J/\psi$, $B\to K^*\psi^\prime$ and
$B\to K^*\gamma$ are all included, and applies only to $a_+$, $f$ and
$g$. On the scale shown, the curves for Fit 4 are indistinguishable
from those of Fit 3. The arrows indicate the
maximum $q^2$ accessible in the semileptonic decays $D\to K^{(*)}\ell
\nu$.\label{formse}}
\end{figure}

\begin{figure}
\centerline{\mbox{\begin{turn}{0}%
\epsfxsize=3.0in\epsffile{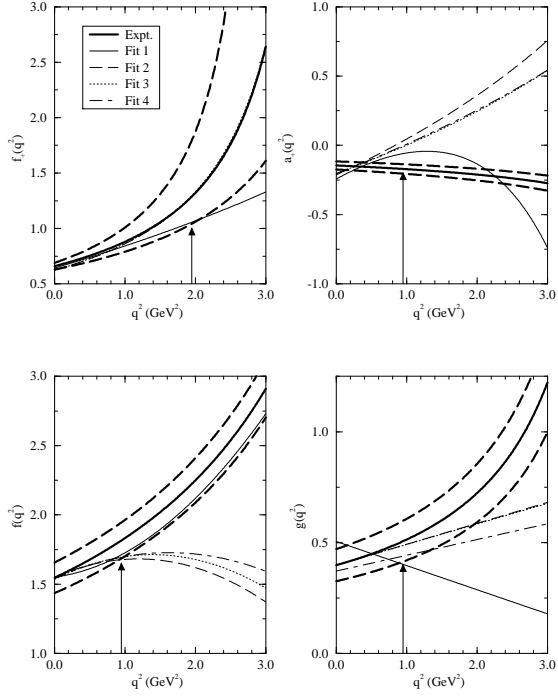}\end{turn}}}
\vskip 0.25in
\caption{Form factors for the $D\to K^{(*)}$ transitions that result
from fits, using `multipolar' forms. The key is as in fig. \protect
\ref{formse}.\label{forms}}
\end{figure}

\begin{figure}
\centerline{\mbox{\begin{turn}{0}%
\epsfxsize=3.0in\epsffile{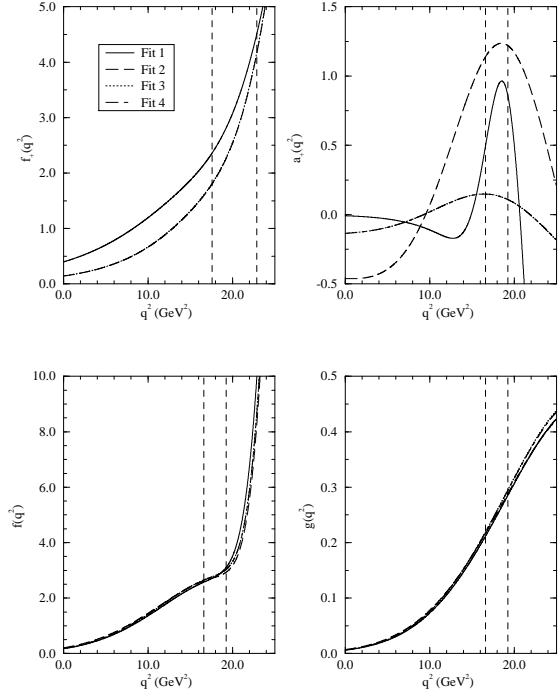}\end{turn}}}
\vskip 0.25in
\caption{Form factors for the $B\to K^{(*)}$ transitions that result
from fits, using exponential forms. In each graph,
Fit 1 means that only $D\to K^{(*)}\ell\nu$ is included in the fit;
Fit 2 means $D\to K^{(*)}\ell\nu$ and $B\to K^{(*)}J/\psi$ are
included; Fit 3 means $D\to K^{(*)}\ell\nu$, $B\to K^{(*)}J/\psi$ and
$B\to K^{(*)}\psi^\prime$ are included; Fit 4
means $D\to K^*\ell\nu$, $B\to K^*J/\psi$, $B\to K^*\psi^\prime$ and
$B\to K^*\gamma$ are all included, and applies only to $a_+$, $f$ and
$g$. The region between the vertical dashed
lines is the range of $q^2$ for which information is available from
the $D\to K^{(*)}\ell\nu$ semileptonic decays. \label{formsbe}}
\end{figure}

\begin{figure}
\centerline{\mbox{\begin{turn}{0}%
\epsfxsize=3.0in\epsffile{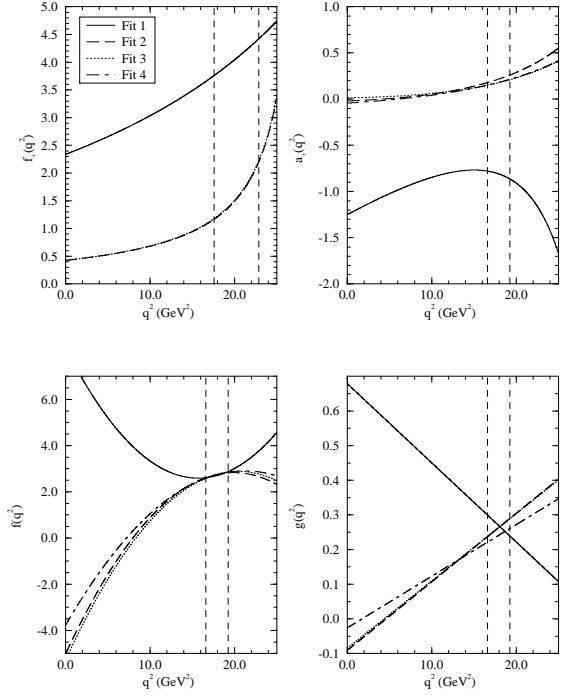}\end{turn}}}
\vskip 0.25in
\caption{Form factors for the $B\to K^{(*)}$ transitions that result
from fits, using multipolar forms. The key is as in fig. \protect
\ref{formsbe}
\label{formsb}}
\end{figure}

The form factors for the $D\to K^{(*)}$ transitions that result from
these fits are shown in fig. \ref{formse} for the exponential forms,
and
in fig. \ref{forms} for the multipolar forms. In each of these
figures, we indicate with arrows the maximum $q^2$ possible in the
corresponding
$D\to K^{(*)}$ decay. The corresponding form factors for the $B\to
K^{(*)}$ transitions are shown in figs. \ref{formsbe} and
\ref{formsb}
respectively. In these latter figures, the vertical dashed lines
indicate the range of $q^2$ that correspond to the $D\to K^{(*)}$
decays: the values
of $v\cdot p$ within the dashed lines of figs \ref{formsbe} and
\ref{formsb} are the same as those that lie to the left of the arrows
in figs. \ref{formse}
and \ref{forms}.

The effect of the mixing mentioned near the end of the previous
section is seen in the curves for $f_+^B$ and $a_+^B$.
In the case of $f_+$ all of the curves for the $D$ form factors are
very close to each other, and are all quite similar to the monopole
form over
the entire range of physically accessible $q^2$. Application of eqn.
(\ref{ffmix}), or more truthfully, of eqn. (\ref{ff1}), leads to
curves for
$f_+^B$ seen in figs. \ref{formsbe} and \ref{formsb}. We point out
that the scaling effect due to the coefficient of the $f_+^D$ term is
only about 15\%, so that the very different forms for $f_+^B$ seen in
the figures must be attributed to the mixing with $f_-^D$.

In examining the curves of this figure, one must remember to compare
the form factors near the kinematic end points. This means
that $f_+^D$ near $q^2=q^2_{{\rm max}_D}=(m_D-m_K)^2=1.95$ GeV$^2$
should be compared with $f_+^B$ near $q^2=(m_B-m_K)^2=24.8$ GeV$^2$.
It is very interesting to note that the form factors for the $B\to
K^{(*)}$ transitions that result from fitting to the semileptonic
decays alone are usually very different from those that result when
the nonleptonic decays are included in the fit.

By examining the graphs of figs. \ref{formse} and \ref{forms}, as
well as the numbers of tables \ref{fitparame1} and \ref{fitparam1},
it is clear that the
form factors that we have obtained are quite consistent with the
experimentally extracted ones, {\it in the range of physically
accessible $q^2$}. Outside of this range,
however, all of the form factors we obtain are
markedly different from the experimental forms. This clearly has very
important consequences
for the nonleptonic and rare decays.

\subsection{Results And Discussion}

As can be seen from the numbers in tables \ref{slfite} and
\ref{slfit}, all fits to the semileptonic decays are reasonable. The
quality of the fit with
respect to the
nonleptonic decays, the results of which are displayed in tables
\ref{nlfite} and \ref{nlfit}, is quite different, however. In the
case of Fit 1, the nonleptonic
decays are generally poorly described, while in the case of Fits 2,
3 and 4, the theory does a reasonable job of describing all of the
data. The differences between Fits 1 and 2 are shown most
graphically
in the form factors of figs. \ref{formsbe} and \ref{formsb}. These
differences have very significant effects on the decay rates for the
rare decays. In general, the differences
in the form factors among Fits 2, 3 and 4 are much less striking. We
also point out that the striking differences in form factors are
achieved, for the most part,
with very small adjustments to the intercepts and slope parameters,
as seen in tables \ref{fitparame1} and \ref{fitparam1}.


\begin{table}
\caption{Results of fits for semileptonic decays, exponential forms.
The columns are as in table \protect\ref{fitparame}.
\label{slfite}}
\begin{tabular}{|l||c|c|c|c|c||}\hline
Quantity & Experiment & Fit 1 & Fit 2 & Fit 3 & Fit 4 \\ \hline
$\Gamma_{D\to K\ell\nu}$ ($10^{-14}$ GeV) & $4.16\pm 0.50$ & $4.26\pm
0.28$ & $3.91\pm 0.24$ & $3.91\pm 0.24$ & - \\ \hline
$\Gamma_{D\to K^*\ell\nu}$ ($10^{-14}$ GeV) & $2.98\pm 0.25$ & $2.88
\pm 0.12$ & $2.87\pm 0.14$ & $2.76\pm 0.12$ & $2.77\pm 0.12$ \\
\hline
$\frac{\Gamma_L}{\Gamma_T}\left(D\to K^*\ell\nu\right)$ & $1.23\pm
0.13 $ & $1.19\pm 0.12$ & $1.23\pm 0.13$ & $1.06\pm 0.08$ & $1.06\pm
0.10$ \\ \hline
$\frac{\Gamma_+}{\Gamma_-}\left(D\to K^*\ell\nu\right)$ & $0.16\pm
0.04$ & $0.20\pm 0.02$ & $0.18\pm 0.02$ & $0.18\pm 0.03$ & $0.19\pm
0.02$ \\ \hline
\end{tabular}
\end{table}

\begin{table}
\caption{Results of fits for semileptonic decays, multipolar forms.
The columns are as in table \protect\ref{fitparame}.
\label{slfit}}
\begin{tabular}{|l||c|c|c|c|c||}\hline
Quantity & Experiment & Fit 1 & Fit 2 & Fit 3 & Fit 4 \\ \hline
$\Gamma_{D\to K\ell\nu}$ ($10^{-14}$ GeV) & $4.16\pm 0.50$ & $3.96\pm
0.28$ & $3.94\pm 0.23$ & $3.95\pm 0.25$ & - \\ \hline
$\Gamma_{D\to K^*\ell\nu}$ ($10^{-14}$ GeV) & $2.98\pm 0.25$ & $2.88
\pm 0.12$ & $2.87\pm 0.14$ & $2.85\pm 0.13$ & $2.88\pm 0.10$ \\
\hline
$\frac{\Gamma_L}{\Gamma_T}\left(D\to K^*\ell\nu\right)$ & $1.23\pm
0.13 $ & $1.21\pm 0.10$ & $1.22\pm 0.09$ & $1.20\pm 0.13$ & $1.23\pm
0.09$ \\ \hline
$\frac{\Gamma_+}{\Gamma_-}\left(D\to K^*\ell\nu\right)$ & $0.16\pm
0.04$ & $0.16\pm 0.03$ & $0.16\pm 0.03$ & $0.17\pm 0.04$ & $0.20\pm
0.03$ \\ \hline
\end{tabular}
\end{table}

\begin{table}
\caption{Results of fits for nonleptonic decays, exponential forms.
The columns are as in table \protect\ref{fitparame}.
\label{nlfite}}
\begin{tabular}{|l||c|c|c|c|c||}\hline
Quantity & Experiment & Fit 1 & Fit 2 & Fit 3 & Fit 4\\ \hline
$\Gamma_{B\to K\psi}$ ($10^{-16}$ GeV) & $3.29\pm 0.95$ & $1.27\pm
15.0\times 10^{-4}$ & $3.29\pm 0.95$ & $3.33\pm 0.95$ & - \\ \hline
$\Gamma_{B\to K^*\psi}$ ($10^{-16}$ GeV) & $6.93\pm 1.33$ & $1.61\pm
1.13$ & $7.06\pm 1.48$ & $6.75\pm 1.11$ & $6.77\pm 1.30$ \\ \hline
$\frac{\Gamma_L}{\Gamma}\left(B\to K^*\psi\right)$ & $0.78\pm 0.073$
& $0.13\pm 0.21$ & $0.77\pm 0.08$ & $0.78\pm 0.05$ & $0.78\pm 0.07$
\\ \hline
$\Gamma_{B\to K\psi^\prime}$ ($10^{-16}$ GeV) & $<3.5$ & $1.57\pm
10.79\times 10^{-3}$ & $1.07\pm 0.25$ & $1.08\pm 0.25$ & - \\ \hline
$\Gamma_{B\to K^*\psi^\prime}$ ($10^{-16}$ GeV) & $6.14\pm 3.95$ &
$22.90\pm 9.10$ & $5.75\pm 2.64$ & $9.46\pm 0.53$ & $9.46\pm 1.18$ \\
\hline
$\frac{\Gamma_L}{\Gamma}\left(B\to K^*\psi^\prime\right)$ & - & $0.07
\pm 0.04$ & $0.96\pm 0.02$ & $0.76\pm 0.03$ & $0.76\pm 0.05$ \\
\hline
\end{tabular}
\end{table}

\begin{table}
\caption{Results of fits for nonleptonic decays, multipolar forms.
The columns are as in table \protect\ref{fitparame}.
\label{nlfit}}
\begin{tabular}{|l||c|c|c|c|c||}\hline
Quantity & Experiment & Fit 1 & Fit 2 & Fit 3 & Fit 4\\ \hline
$\Gamma_{B\to K\psi}$ ($10^{-16}$ GeV) & $3.29\pm 0.95$ & $77.02\pm
59.28$ & $3.79\pm 0.51$ & $3.82\pm 0.65$ & - \\ \hline
$\Gamma_{B\to K^*\psi}$ ($10^{-16}$ GeV) & $6.93\pm 1.33$ & $280.0\pm
52.0$ & $6.96\pm 1.30$ & $6.86\pm 1.34$ & $7.69\pm 1.11$ \\ \hline
$\frac{\Gamma_L}{\Gamma}\left(B\to K^*\psi\right)$ & $0.78\pm 0.073$
& $0.88\pm 0.03$ & $0.78\pm 0.07$ & $0.78\pm 0.07$ & $0.71\pm 0.05$
\\ \hline
$\Gamma_{B\to K\psi^\prime}$ ($10^{-16}$ GeV) & $<3.5$ & $26.76\pm
17.10$ & $1.73\pm 0.33$ & $1.75\pm 0.38$ & - \\ \hline
$\Gamma_{B\to K^*\psi^\prime}$ ($10^{-16}$ GeV) & $6.14\pm 3.95$ &
$26.00\pm 4.50$ & $9.11\pm 0.97$ & $8.33\pm 2.81$ & $8.49\pm 0.80$ \\
\hline
$\frac{\Gamma_L}{\Gamma}\left(B\to K^*\psi^\prime\right)$ & - & $0.69
\pm 0.07$ & $0.71\pm 0.05$ & $0.70\pm 0.08$ & $0.69\pm 0.04$ \\
\hline
\end{tabular}
\end{table}

One comment on the ratio $\rho_L=\Gamma_L/\Gamma$ in $B\to K^*\psi$
is worth making. For many fits, we obtain values for this ratio that
are essentially unity.
Gourdin {\it et al.} \cite{gourdin1,gourdin2} state that, assuming
factorization of the transition amplitude, this ratio has a maximum
value of 0.833. They go
on to point out that observation of a value of $\rho_L$ greater than
this value would be indication of significant non-factorizable
contribution to the
transition amplitude. In their notation,
\begin{equation}
\rho_L=\frac{(a-bx)^2}{(a-bx)^2+2(1+c^2y^2)},
\end{equation}
with $a$, $b$ and $c$ being determined by kinematics, and
\begin{equation}
x=\frac{A_2(m_\psi^2)}{A_1(m_\psi^2)}, \,\,\,\, y=\frac{V(m_
\psi^2)}{A_1(m_\psi^2)}.
\end{equation}
The form factors $A_1$, $A_2$ and $V$ are those defined by Bauer {\it
et al.} \cite{bsw}. The error in their claim arises from ignoring the
possibility that
$x$ may be large, so that $\rho_L\simeq 1$. In our notation, $x
\approx (m_B-m_{K^*})^2a_+/f$. Thus, we would claim that measurement
of $\rho_L$ greater than 0.833
is not an indication that factorization is breaking down. Indeed, we
have assumed factorization, but in fit 1 of table \ref{nlfit}, we
find a value of $\rho_L=0.88
> 0.83$.

\begin{figure}
\centerline{\mbox{\begin{turn}{0}%
\epsfxsize=3.0in\epsffile{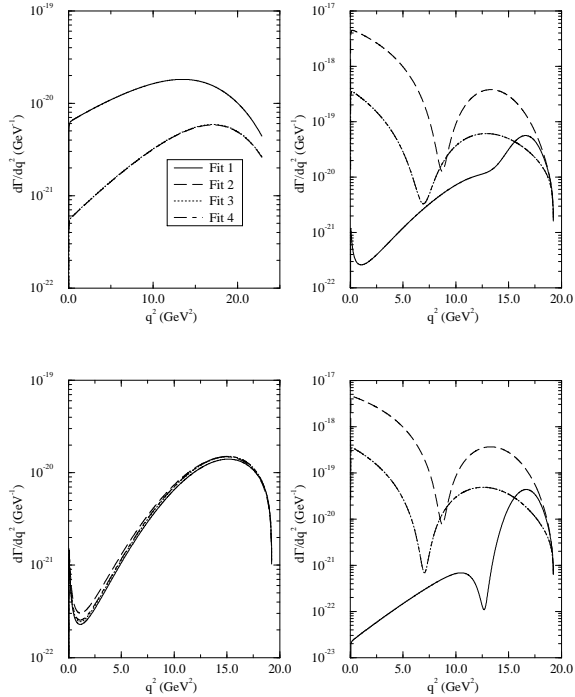}\end{turn}}}
\vskip 0.25in
\caption{Differential decay rates for the processes $B\to K\mu^+
\mu^-$ and $B\to K^*\mu^+\mu^-$, in the exponential
scenario. The graphs are, starting at the top left and moving
clockwise:
$B\to K\mu^+\mu^-$; $B\to K^*\mu^+\mu^-$; $B\to K^*\mu^+\mu^-$ for
longitudinally polarized $K^*$'s; $B\to K^*\mu^+\mu^-$ for
transversely polarized $K^*$'s. In each graph, the key is as in fig.
\protect\ref{formsbe}.
\label{rarerates_e}}
\end{figure}

\begin{figure}
\centerline{\mbox{\begin{turn}{0}%
\epsfxsize=3.0in\epsffile{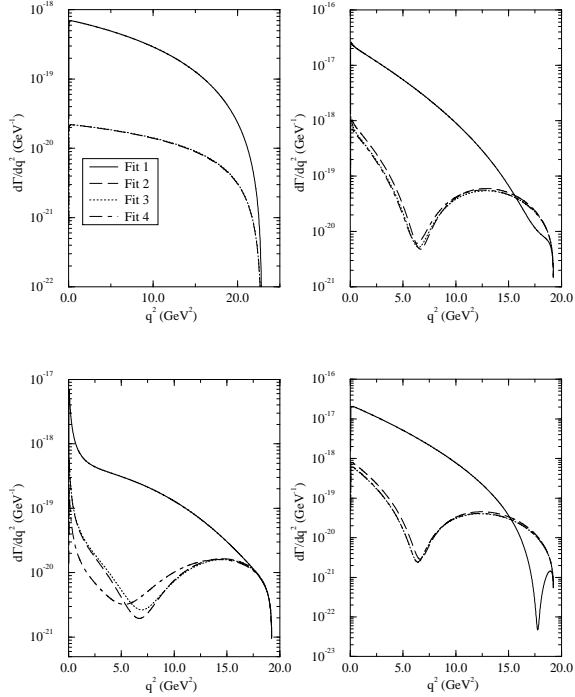}\end{turn}}}
\vskip 0.25in
\caption{Differential decay rates for the processes $B\to K\mu^+
\mu^-$ and $B\to K^*\mu^+\mu^-$, in the multipolar
scenario. The graphs are, starting at the top left and moving
clockwise:
$B\to K\mu^+\mu^-$; $B\to K^*\mu^+\mu^-$; $B\to K^*\mu^+\mu^-$ for
longitudinally polarized $K^*$'s; $B\to K^*\mu^+\mu^-$ for
transversely polarized $K^*$'s.
In each graph, the key is as in fig. \protect\ref{formsbe}.
\label{rarerates}}
\end{figure}

In tables \ref{rfite} and \ref{rfit} we display our predictions for
the rare decays $B\to K^*\gamma$, $B\to K\ell^+\ell^-$ and $B\to K^*
\ell^+\ell^-$, obtained using
the form factors from the four different fits. The lepton spectra are
shown in figs. \ref{rarerates_e} and \ref{rarerates}. As expected,
the form factors from Fit 1,
especially in the exponential scenario, lead to rates that are ruled
out by experimental observations, particularly in the case of the
decay $B\to K^*\gamma$.
In the multipolar scenario, the predictions from Fits 2 and 3 are
somewhat larger than the CLEO \cite{playfer} measurement of $B\to K^*
\gamma$,
while the result from Fit 4 is consistent with
the measured value.

\begin{table}
\caption{Predictions for decay rates of rare processes, exponential
scenario. The columns are as in table \protect\ref{fitparame}.
\label{rfite}}
\begin{tabular}{|l||c|c|c|c|c||}\hline
Quantity & Experiment & Fit 1 & Fit 2 & Fit 3 & Fit 4 \\ \hline
$\Gamma_{B\to K^*\gamma}$ ($10^{-17}$ GeV) & $1.76\pm 0.83$ & $0.01
\pm 0.03$ & $0.02\pm 0.02$ & $0.01\pm 0.01$ & $0.01\pm 0.01$ \\
\hline
$\Gamma_{B\to K\mu^+\mu^-}$ ($10^{-18}$ GeV) & $<158.0$ & $0.30\pm
0.72$ & $0.08\pm 0.02$ & $0.08\pm 0.02$ & - \\ \hline
$\Gamma_{B\to K^*\mu^+\mu^-}^T$ ($10^{-18}$ GeV) & - & $0.12\pm 0.05$
& $0.13\pm 0.03$ & $0.13\pm 0.02$ & $0.13\pm 0.02$ \\ \hline
$\Gamma_{B\to K^*\mu^+\mu^-}^L$ ($10^{-18}$ GeV)  & - & $0.13\pm
0.02$ & $16.3\pm 12.1$ & $1.15\pm 0.20$ & $1.15\pm 1.60$ \\ \hline
$\Gamma_{B\to K^*\mu^+\mu^-}$ ($10^{-18}$ GeV) & $<10.1$ & $0.25\pm
0.06$ & $16.5\pm 12.1$ & $1.27\pm 0.19$ & $1.27\pm 1.60$ \\ \hline
\end{tabular}
\end{table}

\begin{table}
\caption{Predictions for decay rates of rare processes, multipolar
scenario. The columns are as in table \protect\ref{fitparame}.
\label{rfit}}
\begin{tabular}{|l||c|c|c|c|c||}\hline
Quantity & Experiment & Fit 1 & Fit 2 & Fit 3 & Fit 4 \\ \hline
$\Gamma_{B\to K^*\gamma}$ ($10^{-17}$ GeV) & $1.76\pm 0.83$ & $77.90
\pm 17.80$ & $6.17\pm 3.31$ & $6.34\pm 4.35$ & $2.21\pm 0.72$ \\
\hline
$\Gamma_{B\to K\mu^+\mu^-}$ ($10^{-18}$ GeV) & $<158.0$ & $6.33\pm
5.65$ & $0.27\pm 0.05$ & $0.28\pm 0.05$ & - \\ \hline
$\Gamma_{B\to K^*\mu^+\mu^-}^T$ ($10^{-18}$ GeV) & - & $5.05\pm
1.04$ & $0.33\pm 0.13$ & $0.34\pm 0.15$ & $0.22\pm 0.03$ \\ \hline
$\Gamma_{B\to K^*\mu^+\mu^-}^L$ ($10^{-18}$ GeV)  & - & $72.7\pm
14.5$ & $1.82\pm 1.49$ & $1.42\pm 1.25$ & $1.45\pm 0.54$ \\ \hline
$\Gamma_{B\to K^*\mu^+\mu^-}$ ($10^{-18}$ GeV) & $<10.1$ & $77.7\pm
14.4$ & $2.15\pm 1.49$ & $1.76\pm 1.26$ & $1.67\pm 0.55$ \\ \hline
\end{tabular}
\end{table}

Given the failure of the exponential scenario to explain the $B\to
K^*\gamma$ data, one might be tempted to discard its predictions for
the dileptonic
decays. However, we see that the integrated rates are, for the most
part, quite similar to those predicted in the multipolar scenario.
The spectra that result
from the two scenarios are very different, however, and the
predictions for the relative amount of longitudinally polarized and
transversely polarized $K^*$
produced are also somewhat different.
The exponential forms predict $\Gamma_T/
\Gamma_L\approx 0.1$, while in the multipolar scenario, the ratio
ranges between
0.15 and 0.24. The moral here may be that the exponential forms may be
adequate for predicting total rates, but not decay spectra, nor
polarization observables.
We limit our discussion to the multipolar scenario in what follows.

The predictions for the process $B\to K\ell^+
\ell^-$ are two to three orders of magnitude smaller than present
experimental upper limits, while those for $B\to K^*\ell^+\ell^-$ are
smaller than the experimental
limits by factors of four to six. We point out, however,
that the calculated rates for these last two processes do not include
possible contributions from charmonium resonances, which will
certainly alter the
shape of the lepton spectrum, and should also increase the total
decay rate. An investigation of this effect will be left for a future
article. However,
in at least one experimental analysis, kinematic cuts are
imposed on the total mass of the lepton pair, so that events that may
arise from either of the
first two vector charmonium resonances are excluded \cite{playfer}.
In any case, our
results suggest that the exclusive dileptonic decay to the $K^*$
should be observed in the
near future.

The results that we have obtained here again illustrate that the fit
to the present $D\to K^{(*)}$ semileptonic spectra alone is
inadequate for providing information on
$B\to K^*$ processes. Even when we include the nonleptonic decays,
the predictions for different decay modes (particularly $B\to K^*
\ell^+\ell^-$ with longitudinally
polarized $K^*$'s) are sensitive to the nonleptonic modes we include
in the fit. If any of these predictions are to be taken seriously, we
would suggest
that most attention be paid to the predictions of Fit 4 in the
multipolar scenario, as this is the only scenario that adequately
describes all of
the data available.

One of the features of the predicted spectra are the minima in the
differential decay rates. These minima are the result of zeroes in
the respective helicity
amplitudes, and the question of whether or not these zeroes do indeed
exist, and of their exact locations, will have to await a
$B$-factory. However, long-distance
effects, such as those that arise from charmonium resonances, or even
from the charmonium continuum, will at least alter the positions of
the zeroes, and may
wash out the effect altogether.

\section{Conclusion}

We have used the scaling predictions of HQET, together with the most
recent data on $D\to K^{(*)}\ell\nu$ semileptonic decays, to extract
the form factors that describe
the $D\to K^{(*)}$ and $B\to K^{(*)}$ processes. The latter we have
applied to other processes, namely the nonleptonic decays $B\to
K^{(*)}\psi$ and
$B\to K^{(*)}\psi^\prime$, as well as the rare decays $B\to K^*
\gamma$ and $B\to K^{(*)}\ell^+\ell^-$. In the case of the
nonleptonic decays, we have assumed
factorization of the transition amplitude is valid. We have also
performed simultaneous fits of the
semileptonic, nonleptonic and rare processes, and have found that
HQET, together with factorization, provide an adequate framework for
describing the observations.
Our predictions for the modes $B\to K^{(*)}\ell^+\ell^-$ suggest that
these should be measurable in the next generation of experiments, and
certainly at
the proposed $B$ factory.

Perhaps the greatest shortcoming of our fit procedure lies in how we
handle the data, or simulated data, for the semileptonic decays.
Ideally, we should have attempted to
fit our choices of form factors to the experimentally measured
differential decay rates. As a second choice, our choices of form
factors should have been input into
the experimental Monte-Carlo programs to obtain the fit parameters.
In any case, it is clear that the present data, particularly in the
$D\to K^*\ell\nu$ mode, are
inadequate to sufficiently constrain the form factors. In
addition, the differences in form factors between Fit 1 and Fits 2, 3
and 4 are quite striking.

The scenario that best describes all of the experimental data is the
multipolar one, and in this scenario, we find that the universal
form factor $\xi_6$ is linear in $v\cdot p$. Using this scenario,
we predict $Br(\bar B^0\to \bar K^0\mu^+\mu^-)=
(6.4\pm 1.0)\times 10^{-7}$ and $Br(\bar B^0\to \bar K^{*0}\mu^+\mu^-)=
(3.8\pm 1.3)\times 10^{-6}$. These numbers are consistent with other
model calculations \cite{playfer}. We also predict $\Gamma_T/\Gamma_L$ in
$\bar B^0\to \bar K^{*0}\mu^+\mu^-$ to be $0.15\pm 0.07$.

To fully constrain the predictions of HQET, information on the form
factors $a_-$ and $f_-$ is needed from the semileptonic decays. Such
information can only be obtained from high-precision measurements of
the decay spectra at low values of $q^2$, particularly for
semileptonic decays to muons,
as well as by measuring the polarization of the charged lepton, again
preferably the muon. In addition, the precision and statistics in the
$q^2$ spectrum must be improved
so that the form factor parameters can be extracted from the data,
particularly for the decays $D\to K^*\ell\nu$. Perhaps the ideal
experiment would be the equivalent
of present CLEO experiments, in which the machine is tuned to be a
source of $B\bar B$ pairs, produced from the strong decays of the $
\Upsilon(4S)$.
For $D$ decays, the equivalent would be to produce a copious number of
$\psi(3770)$'s, which can be realized at the proposed tau-charm
factory.

\acknowledgements

We gratefully acknowledge helpful conversations with N. Isgur, particularly
for some very useful discussions on scaling relations. We also thank
C. Carlson, A. Freyberger, J. Goity and K. Protasov for discussions.
W. R. acknowledges the support of the National Science
Foundation under grant PHY 9457892, and the U.
S. Department of Energy under contracts DE-AC05-84ER40150 and DE
FG05-94ER40832. W. R. also
acknowledges the hospitality and support of Institut des Sciences
Nucl\'eaires, Grenoble, France, where much of this work was done, and
of Centre
International des Etudiants et Stagiaires.

\end{document}